# Research on Mobile Cloud Computing: Review, Trend and Perspectives

Han Qi
Faculty of Computer Science and Information Technology
University of Malaya
Kuala Lumpur, Malaysia
hanqi@siswa.um.edu.my

Abdullah Gani
Faculty of Computer Science and Information Technology
University of Malaya
Kuala Lumpur, Malaysia
abdullah@um.edu.my

*Abstract*—Mobile Cloud Computing (MCC) which combines mobile computing and cloud computing, has become one of the industry buzz words and a major discussion thread in the IT world since 2009. As MCC is still at the early stage of development, it is necessary to grasp a thorough understanding of the technology in order to point out the direction of future research. With the latter aim, this paper presents a review on the background and principle of MCC, characteristics, recent research work, and future research trends. A brief account on the background of MCC: from mobile computing to cloud computing is presented and then followed with a discussion on characteristics and recent research work. It then analyses the features and infrastructure of mobile cloud computing. The rest of the paper analyses the challenges of mobile cloud computing, summary of some research projects related to this area, and points out promising future research directions.

## I. INTRODUCTION

Over the past few years, advances in the field of network based computing and applications on demand have led to an explosive growth of application models such as cloud computing, software as a service, community network, web store, and so on. As a major application model in the era of the Internet, Cloud Computing has become a significant research topic of the scientific and industrial communities since 2007. Commonly, cloud computing is described as a range of services which are provided by an Internet-based cluster system. Such cluster systems consist of a group of low-cost servers or Personal Computers (PCs), organizing the various resources of the computers according to a certain management strategy, and offering safe, reliable, fast, convenient and transparent services such as data storage, accessing and computing to clients. According to the top ten strategic technology trends for 2012 [1] provided by Gartner (a famous global analytical and consulting company), cloud computing has been on the top of the list, which means cloud computing will have an increased impact on the enterprise and most organizations in 2012.

Meanwhile, smartphones are considered as the representative for the various mobile devices as they have been connected to the Internet with the rapidly growing of wireless network technology. Ubiquity and mobility are two major features in the next generation network which provides a range of personalized network services through numerous network terminals and modes of accessing. The core technology of cloud computing is centralizing computing, services, and specific applications as a utility to be sold like water, gas or electricity to users. Thus, the combination of a ubiquities mobile network and cloud computing generates a new computing mode, namely Mobile Cloud Computing.

As an inheritance and development of cloud computing, resources in mobile cloud computing networks are virtualized and assigned in a group of numerous distributed computers rather than in traditional local computers or servers, and are provided to mobile devices such as smartphones, portable terminal, and so on. (see Fig. 1). Meanwhile, various applications based on mobile cloud computing have been developed and served to users, such as Googles Gmail, Maps and Navigation systems for Mobile, Voice Search, and some applications on an Android platform, MobileMe from Apple, Live Mesh from Microsoft, and MotoBlur from Motorola. According to the research from Juniper, the cloud computing based mobile software and application are expected to rise 88% annually from 2009 to 2014, and such growth may create US 9.5 billion dollars in 2014.

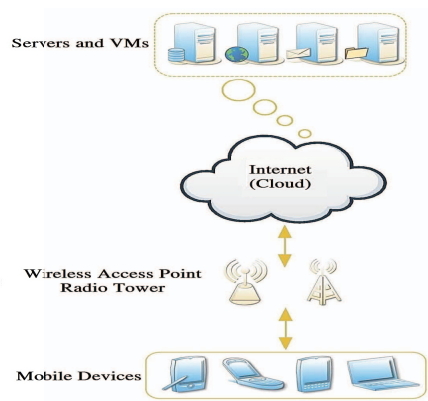

Fig. 1: Mobile Cloud Computing

While mobile cloud computing make a great contribution to our daily lives, it will also, however, bring numerous challenges and problems. In short, the core of such challenges and problems is just how to combine the two technologies seamlessly. On one hand, to ensure that mobile devices adequately make best use of advantages of cloud computing to improve and extend their functions. On the other hand, to overcome the disadvantages of limited resources and computing ability in mobile devices in order to access cloud computing with high efficiency like traditional PCs and Servers. Thus, in order to solve the mentioned challenges and point out further research, getting a thorough understanding of the novel computing paradigm - mobile cloud computing, is necessary. This paper introduces the basic model of mobile cloud computing, its background, key technology, current research status, and its further research perspectives as well.

## II. BACKGROUND

As a development and extension of Cloud Computing and Mobile Computing, Mobile Cloud Computing, as a new phrase, has been devised since 2009. In order to help us grasping better understanding of Mobile Cloud Computing, let's start from the two previous techniques: Mobile Computing and Cloud Computing.

### A. Mobile Computing

Mobility has become a very popular word and rapidly increasing part in today's computing area. An incredible growth has appeared in the development of mobile devices such as, smartphone, PDA, GPS Navigation and laptops with a variety of mobile computing, networking and security technologies. In addition, with the development of wireless technology like WiMax, Ad Hoc Network and WIFI, users may be surfing the Internet much easier but not limited by the cables as before. Thus, those mobile devices have been accepted by more and more people as their first choice of working and entertainment in their daily lives.

So, what is Mobile computing exactly? In Wikipedia, it is described as a form of human-computer interaction by which a computer is expected to be transported during normal usage [2]. Mobile computing is based on a collection of three major concepts: hardware, software and communication. The concepts of hardware can be considered as mobile devices, such as smartphone and laptop, or their mobile components. Software of mobile computing is the numerous mobile applications in the devices, such as the mobile browser, anti-virus software and games. The communication issue includes the infrastructure of mobile networks, protocols and data delivery in their use. They must be transparent to end users.

*1) Features:* the features of mobile computing are as follows:

  *a) mobility:* mobile nodes in mobile computing network can establish connection with others, even fixed nodes in wired network through Mobile Support Station (MSS) during their moving.

  *b) Diversity of network conditions:* normally the networks using by mobile nodes are not unique, such networks can be a wired network with high-bandwidth, or a wireless Wide Area Network (WWAN) with low-bandwidth, or even in status of disconnected.

  *c) Frequent disconnection and consistency:* as the limitation of battery power, charge of wireless communication, network conditions and so on, mobile nodes will not always keep the connection, but disconnect and consistent with the wireless network passively or actively.

  *d) Dis-symmetrical network communication:* servers and access points and other MSS enable a strong send/receive ability, while such ability in mobile nodes is quite weak comparatively. Thus, the communication bandwidth and overhead between downlink and uplink are discrepancy.

  *e) Low reliability:* due to signals is susceptible to interference and snooping, a mobile computing network system has to be considered from terminals, networks, database platforms, as well as applications development to address the security issue.

*2) Challenges:* Compared with the traditional wired network, mobile computing network may face various problems and challenges in different aspects, such as signal disturbance, security, hand-off delay, limited power, low computing ability, and so on. due to the wireless environment and numerous mobile nodes. In addition, the Quality of Service (QoS) in mobile computing network is much easier to be affected by the landforms, weather and buildings.

### B. Cloud Computing

In the era of PC, many users found that the PCs they bought 2 years ago cannot keep pace with the development of software nowadays; they need a higher speed CPU, a larger capacity hard disk, and a higher performance Operation System (OS). That is the magic of 'Moores Law' which urges user upgrading their PCs constantly, but never ever overtaken the development of techniques. Thus, a term called 'Cloud Computing' burst upon our lives.

Cloud Computing has become a popular phrase since 2007. However, there is no consensual definition on what a Cloud Computing or Cloud Computing System is, due to dozens of developers and organizations described it from different perspectives. C. Hewitt [3] introduces that the major function of a cloud computing system is storing data on the cloud servers, and uses of cache memory technology in the client to fetch the data. Those clients can be PCs, laptops, smartphones and so on. R. Buyya [4] gives a definition from the perspective of marking that cloud computing is a parallel and distributed computing system, which is combined by a group of virtual machines with internal links. Such systems dynamically offer computing resources from service providers to customers according to their Service level Agreement (SLA). However, some authors mentioned that cloud computing was not a completely new concept. L. Youseff [5] from UCSB argue that cloud computing is just combined by many existent and few new concepts in many research fields, such as distributed and grid computing, Service-Oriented Architectures (SOA) and in virtualization.

In this paper, we consider the cloud computing is a large-scale economic and business computing paradigm with virtualization as its core technology. The cloud computing system is the development of parallel processing, distributed and grid computing on the Internet, which provides various QoS guaranteed services such as hardware, infrastructure, platform, software and storage to different Internet applications and users.

*1) Framework:* cloud computing systems actually can be considered as a collection of different services, thus the framework of cloud computing is divided into three layers, which are infrastructure layer, platform layer, and application layer (see Fig. 2).

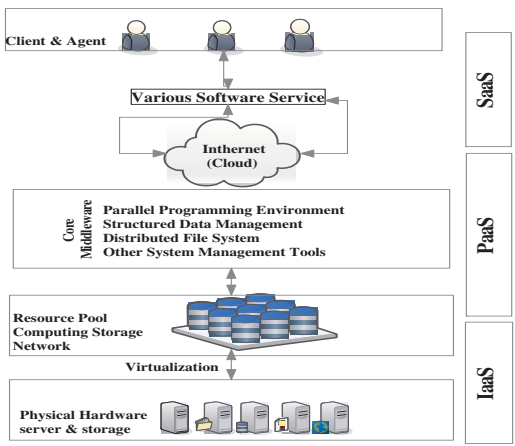

Fig. 2: The Framework of Cloud Computing

*a) Infrastructure layer:* it includes resources of computing and storage. In the bottom layer of the framework, physical devices and hardware, such as servers and storages are virtualized as a resource pool to provide computing storage and network services users, in order to install operation system (OS) and operate software application. Thus it is denoted as Infrastructure as a Service (IaaS). Typically services in this layer such as Elastic Computing Cloud of Amazon [6].

*b) Platform layer:* this layer is considered as a core layer in the cloud computing system, which includes the environment of parallel programming design, distributed storage and management system for structured mass data, distributed file system for mass data, and other system management tools for cloud computing. Program developers are the major clients of the platform layer. All platform resources such as program testing, running and maintaining are provided by the platform directly but not to end users. Thus, this type of services in a platform layer is called Platform as a Service (PaaS). The typical services are Google App Engine [7] and Azure from Microsoft [8].

*c) Application layer:* this layer provides some simple software and applications, as well as costumer interfaces to end users. Thus we name this type of services in the application layer as Software as a Service (SaaS). Users use client software or a browser to call services from providers through the Internet, and pay costs according to the utility business model (like water or electricity) [9]. The earliest SaaS is the Customer Relationship Management (CRM) [10] from Salesforce, which was developed based on the force.com (a PaaS in Salesforce). Some other services provided by Google on-line office such as documents, spreadsheets, presentations are all SaaS.

*2) Features:* the features of Cloud Computing are as follows:

*a) Virtualization:* the 'Cloud' can be considered as a virtual resource pool [11] where all bottom layer hardware devices is virtualized. End users access desired resources through a browser and get data from cloud computing providers without maintaining their own data centres. Furthermore, some virtual machines (VMs) are often installed in a server in order to improve the efficiency to use resources; and such VMs support load migration when there is a server over-load.

*b) Reliability, usability and extensibility:* cloud computing provides a safe mode to store user's data while users do not worry about the issues such as software updating, leak patching, virus attacks and data loss. If failure happens on a server or VM, the cloud computing systems transfer and backup those data to other machines, and then delete those failure nodes from the systems automatically in order to make sure the whole system has normal operation [12]. Meanwhile, cloud can be extended from horizontal and vertical [13] in a large-scale network, to process numerous requests from thousands of nodes and hosts.

*c) Large-scale:* in order to possess the capability of supercomputing and mass storage, a cloud computing system normally consists of thousands of servers and PCs. Google Cloud Computing, for example, has already controlled 2% of all servers or about 1 million servers located in two hundred different places in the world, and will move upward to 10 million servers in the next decade [14].

*d) Autonomy:* a cloud system is an autonomic system, which automatically configures and allocates the resources of hardware, software and storage to clients on-demand, and the management is transparent to end users.

*3) Challenges:* first of all, cloud computing needs an improved mechanism to provide a safe and high efficiency service as the numerous invoked third-party software and infrastructures are implementing in computing. In addition, due to data centres of resource using a mass of electricity, efficient resource scheduling strategy and methods are required in order to save energy. Furthermore, as a Service Level Agreement (SLA) is established between users and service providers in cloud computing, so the performance and analysis of services are necessary to be monitored. Last but not least, simple and convenient application interfaces are indispensable for service providers in cloud computing, thus a uniform standard is required eagerly.

## III. MOBILE CLOUD COMPUTING

Nowadays, both hardware and software of mobile devices get greater improvement than before, some smartphones such as iPhone 4S, Android serials, Windows Mobile serials and Blackberry, are no longer just traditional mobile phones with conversation, SMS, Email and website browser, but are daily necessities to users. Meanwhile, those smartphones include various sensing modules like navigation, optics, gravity, orientation, and so on. which brings a convenient and intelligent mobile experience to users. In 2010, Google CEO Eric Schmidt described mobile cloud computing in an interview that 'based on cloud computing service development, mobile phones will become increasingly complicated, and evolve to a portable super computer' [15]. In the face of various mobile cloud services provided by Microsoft, Apple, Google, HTC, and so on, users may be confused about what mobile cloud computing exactly is, and what its features are.

### A. Concept and principle

Similar with Cloud Computing, there are a lot but no consensual definitions on what mobile cloud computing is. In this paper, we consider it is a novel computing mode consisting of mobile computing and cloud computing, which provide cloud based services to users through the Internet and mobile devices. On one hand, the mobile cloud computing is a development of mobile computing, and an extension to cloud computing. In mobile cloud computing, the previous mobile device-based intensive computing, data storage and mass information processing have been transferred to 'cloud' and thus the requirements of mobile devices in computing capability and resources have been reduced, so the developing, running, deploying and using mode of mobile applications have been totally changed. On the other hand, the terminals which people used to access and acquire cloud services are suitable for mobile devices like smartphone, PDA, Tablet, and iPad but not restricted to fixed devices (such as PC), which reflects the advantages and original intention of cloud computing. Therefore, from both aspects of mobile computing and cloud computing, the mobile cloud computing is a combination of the two technologies, a development of distributed, grid and centralized algorithms, and have broad prospects for application.

As shown is the Fig. 3, mobile cloud computing can be simply divided into cloud computing and mobile computing. Those mobile devices can be laptops, PDA, smartphones, and so on. which connects with a hotspot or base station by 3G, WIFI, or GPRS. As the computing and major data processing phases have been migrated to 'cloud', the capability requirement of mobile devices is limited, some low-cost mobile devices or even non-smartphones can also achieve mobile cloud computing by using a cross-platform mid-ware. Although the client in mobile cloud computing is changed from PCs or fixed machines to mobile devices, the main concept is still cloud computing. Mobile users send service requests to the cloud through a web browser or desktop application, then the management component of cloud allocates resources to the request to establish connection, while the monitoring and calculating functions of mobile cloud computing will be implemented to ensure the QoS until the connection is completed.

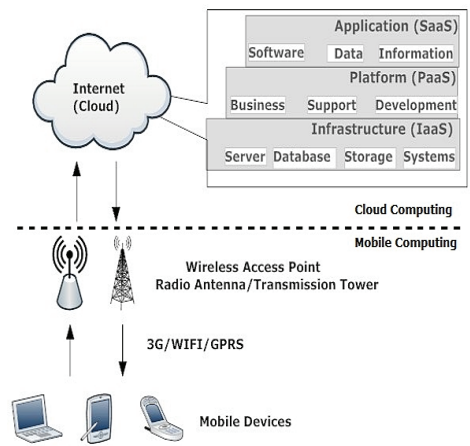

Fig. 3: Architecture of Mobile Cloud Computing

### B. Challenges and solutions

The main objective of mobile cloud computing is to provide a convenient and rapid method for users to access and receive data from the cloud, such convenient and rapid method means accessing cloud computing resources effectively by using mobile devices. The major challenge of mobile cloud computing comes from the characters of mobile devices and wireless networks, as well as their own restriction and limitation, and such challenge makes application designing, programming and deploying on mobile and distributed devices more complicated than on the fixed cloud devices [16]. In mobile cloud computing environment, the limitations of mobile devices, quality of wireless communication, types of application, and support from cloud computing to mobile are all important factors that affect assessing from cloud computing. Table 2 gives an overview of proposed challenges and some solutions about mobile cloud computing.

*1) Limitations of mobile devices:* While discussing mobile devices in cloud the first thing is resource-constrain. Though smartphones have been improved obviously in various aspects such as capability of CPU and memory, storage, size of screen, wireless communication, sensing technology, and operation systems, still have serious limitations such as limited computing capability and energy resource, to deploy complicated applications. By contrast with PCs and Laptops in a given condition, these smartphones like iPhone 4S, Android serials, Windows Mobile serials decrease 3 times in processing capacity, 8 times in memory, 5 to 10 times in storage capacity and 10 times in network bandwidth.

Normally, smartphone needs to be charged everyday as dialling calls, sending messages, surfing the Internet, community accessing, and other internet applications. According to past development trends, the increased mobile computing ability and rapid development of screen technology will lead to more and more complicated applications deployed in smartphones. If the battery technology cannot be improved in a short time, then how to effectively save battery power in smartphone is a major issue we meet today.

TABLE I: Challenges and Solutions of Mobile Cloud Computing

| Challenges | Solutions |
|---|---|
| Limitations of mobile devices | Virtualization and Image, Task migration |
| Quality of communication | Bandwidth upgrading, Data delivery time reducing |
| Division of applications services | Elastic application division mechanism |

The processing capacity, storage, battery time, and communication of those smartphones will be improved consistently with the development of mobile computing. However, such enormous variations will persist as one of major challenges in mobile cloud computing.

*2) Quality of communication:* In contrast with wired network uses physical connection to ensure bandwidth consistency, the data transfer rate in mobile cloud computing environment is constantly changing and the connection is discontinuous due to the existing clearance in network overlay. Furthermore, data centre in large enterprise and resource in Internet service provider normally is far away to end users, especially to mobile device users. In wireless network, the network latency delay may 200 ms in 'last mile' but only 50 ms in traditional wired network.

Some other issues such as dynamic changing of application throughput, mobility of users, and even weather will lead to changes in bandwidth and network overlay. Therefore, the handover delay in mobile network is higher than in wired network.

*3) Division of application services:* In mobile cloud computing environment, due to the issue of limited resources, some applications of compute-intensive and data-intensive cannot be deployed in mobile devices, or they may consume massive energy resources. Therefore, we have to divide the applications and use the capacity of cloud computing to achieve those purposes, which is: the core computing task is processed by cloud, and those mobile devices are responsible for some simple tasks only. In this processing, the major issues affecting performance of mobile cloud computing are: data processing in data centre and mobile device, network handover delay, and data delivery time.

For a given standard, providing a quality guaranteed cloud service should consider the following facts: optimal division of application between cloud and mobile device, interaction between low-latency and code offload, high-bandwidth between cloud and mobile device for high speed data transmission, user-oriented cloud application performance, self-adaptation mechanism of mobile cloud computing, and optimal consumption and overhead of mobile devices and cloud servers. The following strategies can be used to response to the challenges:

1. Upgrade bandwidth for wireless connection, make the web content more suitable for mobile network using regional data centres.
2. Deploy the application processing node at the 'edge' of cloud in order to reduce data delivery time.
3. Duplicate mobile devices to cloud using virtualization and image technologies, to process Data-Intensive Computing (DIC) and Energy-Intensive Computing, such as virus scanning in mobile devices.
4. Dynamically optimize application push in cloud and the division with mobile terminals.

*C. Related work*

So far, industrial and scientific communities have been doing various researches for responding to the above challenges. Some typical research projects and cases are presented in the following.

*1) Augmented Execution:* So far, industrial and scientific communities have been doing various researches for responding to the above challenges. Some typical research projects and cases are presented in the following.

CloneCloud is introduced by B. Chun [17] in 2011. The core method is using virtual machine migration technology to offload execution blocks of applications from mobile devices to Clone Cloud seamlessly and partly, in order to fully or semi-automatically extend or modify the smartphone-based execution to a distributed environment (smartphone computing plus cloud computing). In a CloneCloud system (see Fig. 4),

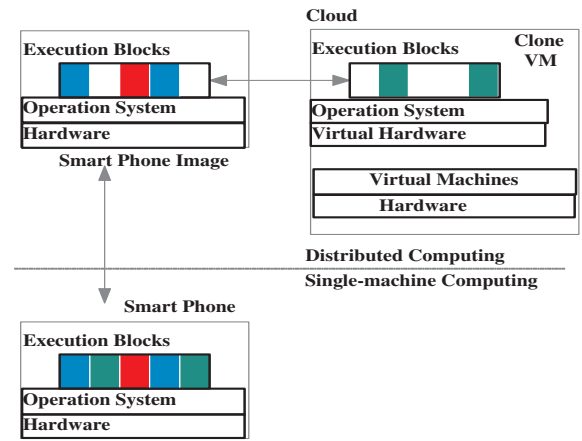

Fig. 4: CloueCloud System Architecture

the 'Clone' is a mirror image of a smartphone running on a virtual machine. By contrast with smartphones, such a 'clone' has more hardware, software, network, energy resources in a virtual machine which provides more suitable environment to process complicated tasks. In the diagram, a task in smartphone is divided to 5 different execution blocks (we mark them as different colors), and the smartphone is cloned (virtualized) as an image in distributed computing environment. Then the image passes some computing or energy-intensive blocks (the Green blocks) to cloud for processing. Once those execution blocks have been completed, the output will be passed from

CloneCloud to the smartphone. A major advantage of the CloneCloud is enhanced smartphones performance. Byung takes a test by implementing a face tracking application in a smartphone with and without CloneCloud. The result shows that only 1 second is spent in CloneCloud environment but almost 100 seconds in the smartphone without CloneCloud. Another advantage of CloneCloud is reduced battery consumption as smartphones do not use its CPU as frequently. The disadvantages of CloneCloud are handover delay, bandwidth limitation. As we know that the speed of data transmission between smartphones and base station is not consistent (according to the situation), therefore, the CloneCloud will be unavailable if mobile users walk in the signal's blind zone.

Based on the CloneCloud, X. Zhang has introduced an Elastic application programming model for mobile cloud computing in [18] to remove the constraints of mobile platforms by extending these mobile terminals to cloud through a distributed framework. This model divides a single application into a range of elasticity patterns called weblets, and dynamic adaptation of configuration running on internet-based cloud and mobile devices. Thus, the capability of mobile device can be enhanced to process for more comprehensive tasks. Furthermore, a cost model is applied in Zhangs research to adjust the patterns execution configurations. However, in this model, we still need a mechanism to manage the communication between weblets in mobile devices due to such devices changing their communication channel (such as 3G to WIFI or GPRS). Another challenge for this model is that a high-speed bandwidth or media channel is required to ensure the quality of communication between weblets.

Although the above methods can reduce power consumption on smartphones effectively, they may still meet a potential long interaction response in data transmission between a cloud and terminals. Therefore, offloading all applications from smartphones to the cloud cannot be justified for power consumption, especially for some lightweight applications which are suitable to be deployed in local smartphones. (But is not worth to be delivered to cloud). Y. Lu [19] proposed a solution, called Virtualized Screen, to move screen rendering from smartphones to a cloud as a service. In his method, only part of smartphone's screen is virtualized in cloud, which involves a collection of data in display images, text contents, video and audio, input of keyboard, touching, and pen on smartphones. The other applications with energy-intensive computing run on cloud. Therefore, parts of applications and interactions are offloaded and executed in cloud, and some light power consumption operation or applications are deployed in local smartphones, which could effectively reduce power consumption and interaction delay. There still remains a future research topic in this area: creating an optimal mechanism to decide which application is deployed in cloud, and which one in local smartphones. In addition, some other issues such as privacy, security or trustworthiness also need to be considered in the migration process.

*2) Elastic Applications:* In order to provide a more effectively mobile cloud application, researchers have developed and extended CloneCloud-based algorithms using dynamically migrating partition of applications to the remote server in cloud.

AlfredO [20] is a middleware platform to automatically distribute different layers of application in smartphones and cloud, respectively, by modelling applications as a consumption graph, and finding the optimal modules. The test result shows that such platform improves the performance of applications in cloud computing effectively. AlfredO system consists of three bundles (the interface encapsulation on Java classes and services): AlfredOClient and Renderer on the client and AlfredOCore on the server (shown in Fig. 5).

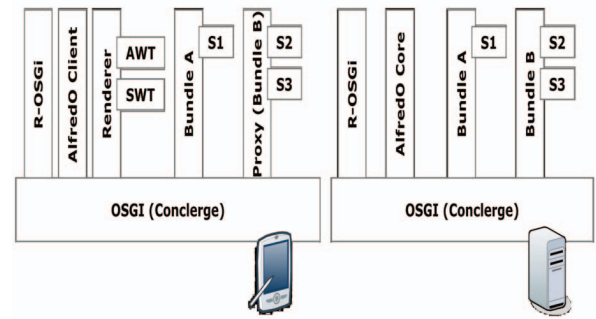

Fig. 5: AlfredO Architecture

When a client requests an application, AlfredOCore first models such application and computes the optimal deployment, and then send application descriptor and the list of services to AlfredO Client. Renderer uses the descriptor to generate the corresponding AWT or SWT interface, while AlfredO Client fetches the specified services via R-OSGi [21].Similar with [19], AlfredO executes parts of application remotely to save battery energy and extend resource of the mobile device effectively. However, such models do not support platform-independent cooperative interaction in an open network, and this issue is needed to be considered in future research.

S. Jeong [22] from Samsung introduces a novel elastic application model which provides a seamless and transparent use of cloud to extend and solve the limitation of mobile devices. This model enables a partition to a single application into multiple components called Weblet, and dynamically deploys these Weblets in execution according to a configuration strategy at cloud and mobile terminals. However, some overhead is generated in the communication among Weblets, between the Internet and Weblets, and the implementing Weblets during the model processing. In order to minimise the above extra overhead and optimise the cost of elastic applications, the authors presented a cost model in their framework, which collects sensor data (such as battery life, loads of devices and cloud, network conditions and so on.) from both mobile devices and cloud as input, and implements the optimal algorithm to dynamically output an execution configuration for the applications, such as deployment of Weblet, resource allocating of cloud, selecting of different network connection, and so on.

*3) Migration Optimization:* As the mobility feature in mobile devices, provide a seamless migration environment for data transmission or service guarantee has becoming another hot issue in mobile cloud computing research. An optimal migration mechanism can reduce interaction delay, enhance processing capability, and improve user's experience effectively.

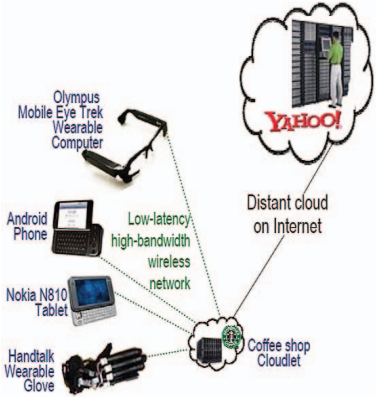

Fig. 6: Concept and Infrastructure of Cloudlet [23]

Cloudlet is presented by M. Satyanarayanan [23] from Carnegie Mellon University, which provides rapidly instantaneous customized service to mobile devices using virtual machine (VM) technology for solving bandwidth-induced delay between devices and cloud, and so on. The author argued that although cloud computing is a good solution to mobile device's resource-constraint, long WAN latent delay is an obstacle for its performance. In a mobile cloud computing environment, the different accessing bandwidth between mobile devices and cloud may lead to different scales of delay, especially when mass data is being transferred and processed, users do feel the existence of such delays. Unfortunately, some kind of delay, such as data checking or firewall filtering for security is inevitable. Therefore, the authors deployed Cloudlet as a 'Micro Cloud' to be accessed by mobile devices with high bandwidth and low delay. Fig. 6 shows that mobile devices use WIFI or WLAN to access Cloudlet which is located in a coffee shop, and then rapidly provides customized service using VM technology.

As for the features of resource-constrains in mobile devices, many researchers are seeking how to solve it. Hyrax is a system developed by E. Marinelli [24] from Carnegie-Mellon University, which deploys Android-based (an open source system) mobile phones as nodes to create a mobile cloud computing platform. This system transplants a modified Hadoop (a framework of cloud from Apache) into Android so that these smartphones can be like PCs to deploy a real cloud computing system. The infrastructure of Hyrax is shown as Fig. 7.

In order to improve the whole performance of Hyrax, smartphone acts as Slave in Hadoop network, but Master is still deployed on PC, NameNode and JobTracker are implemented as background services, and Hadoop Distributed File System (HDFS) is used to store data. In order to evaluate the performance of Hyrax, authors took a lot of tests in Sort, Random Writer, Pi Estimator, Grep, and Word Count using 10 Android G1 phones and 5 HTC Magic phones. The result shows that the performance of smartphones is much worse which took 15 times as long as PC in Map and Reduce procedures. As the first mobile phone based cloud computing system, Hyrax argued that the feature of resource-constraints in mobile phones is the main reason affecting cloud performance, and it also indicated the direction for further research.

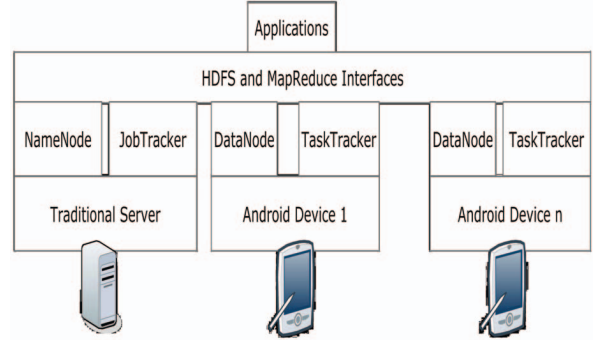

Fig. 7: Hyrax Infrastructure

## IV. OPEN RESEARCH ISSUES

Although some projects of mobile cloud computing have already been deployed around the world, there is still a long way for business implementation, and some research aspects should be considered in further work.

*A. Data delivery*

Due to the feature of resource-constrains, mobile devices have potential challenges in cloud accessing, consistent accessing, data transmission, and so on. Such challenges can be solved using: special application (service) and middle-ware (provide a platform for all mobile cloud computing systems).

*B. Task division*

Researchers divide tasks (applications) from mobile devices into multiple sub-tasks and deliver some of them to run in cloud, which is a good solution to the resource limited mobile devices. However, we do not have an optimal strategy or algorithm on how to divide these tasks, which one should be processed by cloud and which one by devices.

*C. Better service*

The original purpose of mobile cloud computing is providing PC-liked services to mobile terminals. However, as the existing different features between mobile devices and PCs, we cannot directly transplant the services from PCs' platform to mobile devices. Therefore, further research should try to identify the method on how to provide suitable and friendly interactive services for mobile devices.

## V. Conclusion

With the high increasing of data computation in commerce and science, the capacity of data processing has been considered as a strategic resource in many countries. Mobile cloud computing (MCC), as a development and extension of mobile computing (MC) and cloud computing (CC), has inherited the high mobility and scalability, and become a hot research topic in recent years. We conclude that there are three main optimization approaches in MCC, which are focusing on the limitations of mobile devices, quality of communication, and division of applications services. Firstly, using virtualization and image technology can address it effectively, and immigrate task from terminal to cloud is also a good way to achieve better results. Secondly, as we know the quality of communication in wired network is better than in wireless network, so reducing the proportion of data delivery in wireless environment is an effective way to improve the quality. In addition, upgrading bandwidth is envisaged to be a simple way to increase performance but it incurs additional cost to users. Deploying an effective elastic application division mechanism is deemed to be the best solution to guarantee the application service in MCC; its complicated, but promising high impact results.

## Acknowledgment

This work is fully funded by the Malaysian Ministry of Higher Education under the University of Malaya High Impact Research Grant UM.C/HIR/MOHE/FCSIT/03.